\documentclass[12pt]{article} \textwidth=6in \oddsidemargin=0in
\usepackage{amssymb} 
\begin{document}
\def\ov{\over} \def\be{\begin{equation}} 
\def\ee{\end{equation}} \def\R{\mathbb R} 
\def\ph{\varphi} \def\inv{^{-1}} \def\x{\xi} \def\d{\delta} 
\def\b{\beta} \def\a{\alpha} \def\th{\theta} 
\def\bc{\begin{center}} \def\ec{\end{center}} \def\iy{\infty}
\def\ps{\psi} \def\ld{\ldots} \def\g{\gamma} \def\T{\mathbb T}
\def\cd{\cdots} \def\ep{\varepsilon} \renewcommand{\Re}{\mathcal Re} \renewcommand{\Im}{\mathcal Im} \def\O{\Omega}
\def\g{\gamma} \def\e{\eta} \def\noi{\noindent} 
\def\D{\Delta} \def\qed{\hfill$\Box$} \def\sp{\vspace{.5ex}}
\newcommand{\ch}{\raisebox{.2ex}{$\chi$}} \def\m{\mu} 
\def\Del{\nabla} \def\Z{\mathbb Z} \def\s{\sigma} 
\def\M{\mathcal M} \def\C{\mathcal C} \def\z{\zeta} 
\def\hf{{1\ov2}} \def\la{\langle} \def\ra{\rangle} \def\U{\mathcal U}

\hfill March 16, 2014
\bc{\large\bf On the Singularities in the Susceptibility Expansion\\
\vskip1ex 
for the Two-Dimensional Ising Model}\ec

\bc{\large\bf Craig A.~Tracy}\\
{\it Department of Mathematics \\
University of California\\
Davis, CA 95616, USA}\ec

\bc{\large \bf Harold Widom}\\
{\it Department of Mathematics\\
University of California\\
Santa Cruz, CA 95064, USA}\ec

\begin{abstract}
For temperatures below the critical temperature, the magnetic susceptibility for the two-dimensional isotropic Ising model can be expressed in terms of an infinite series of multiple integrals. With respect to a parameter related to temperature and the interaction constant, the integrals may be extended to  functions analytic outside the unit circle. In a groundbreaking paper, 
B.~G.~Nickel \cite{nickel1} identified a class of singularities of these integrals on the unit circle. In this note we show that there are no other singularities on the unit circle.
\end{abstract}

\bc{\bf I. Introduction}\ec

For the two-dimensional zero-field Ising model on a square lattice,  the magnetic susceptibility  as a function of temperature is usually studied through its relation with the zero-field spin-spin correlation function:
\be \b\inv \ch = \sum_{M,N\in\Z}\left\{\la \s_{0,0}\,\s_{M,N} \ra-\M^2 \right\} \label{chi}\ee
where $\b=(k_B T)^{-1}$, $T$ is temperature, $k_B$ is Boltzmann's constant and  $\M$ is the spontaneous magnetization (see, e.g., \cite{mccoy}).
Fisher \cite{fisher} in 1959 initiated the analysis of the analytic structure of $\ch$ near the critical temperature $T_c$ by relating it to the long-distance asymptotics of the correlation
function at $T_c$ (a result known to Kaufman and Onsager).  
Subsequently Wu \textit{et al.}\ \cite{WMTB} derived the exact \textit{form factor expansion} of $\ch$ which has the structure of an infinite series whose $n$th  order term is an $n$-dimensional integral. In later work \cite{PT,  Ya, palmer} the structure of 
the integrands of these $n$-dimensional integrals was simplified.  

The analysis of $\ch$ as a function of the \textit{complex variable} $T$ was initiated by Guttmann and Enting \cite{GE} where, by the use of high-temperature series expansions, they were led to conjecture that $\ch$, as a function of $T$, possesses a natural boundary.  In two groundbreaking papers, Nickel \cite{nickel1, nickel2}  analyzed the $n$-dimensional integrals appearing in the form factor expansion of $\ch$ and identified a class of complex singularities, now called \textit{Nickel singularities}, that lie on a curve and which become ever more dense with increasing $n$.  This work of Nickel provides  very strong support for the existence of a natural boundary for $\ch$.\footnote{As Nickel noted, for a rigorous proof one must show that there are no cancellations of the singularities in the infinite sum.}  For further developments see Chan \textit{et al.}\ \cite{chan} and the review article \cite{mccoy2}.

We  recall that if $T_c$ denotes the critical temperature, then for the isotropic Ising model, where horizontal and vertical interaction constants have the same value~$J$,
the spontaneous magnetization is given for $T<T_c$ by \cite{yang, mccoy}
\[ \M=(1-k^2)^{1/8},\]
where $k:=\left(\sinh 2\beta J\right)^{-2}$; and $\M$ is zero for $T>T_c$. Thus $k=1$ defines the critical temperature $T_c$ and $0<k<1$ is the
region $0<T<T_c$.
Boukraa \textit{et al.}\ \cite{bhmmz} (building on work of Lyberg and McCoy \cite{LM}) introduced a simplified model for $\ch$, called
 the \textit{diagonal susceptibility} $\ch_d$ which has the following analogous representation to (\ref{chi}):
 \[ \beta^{-1} \ch_d = \sum_{N\in\Z}\left\{ \langle \sigma_{0,0} \sigma_{N,N}\rangle-\M^2\right\}. \]
By an analysis similar to that of Nickel, they are led to conjecture a natural boundary for $\ch_d$; which in terms of the complex variable $k$,
 is the unit circle $\vert k \vert=1$.   This conjecture thus says that the low temperature phase, $T<T_c$, is separated from the high-temperature phase $T>T_c$ by the
natural boundary $\vert k \vert =1 $.   This conjecture for $\ch_d$ is precisely the same as the conjectured natural boundary for $\ch$.
In the low-temperature phase, the present authors proved that $\vert k \vert =1$ is a natural boundary 
for $\ch_d$ \cite{tw} thus adding additional support for the conjecture for $\ch$.

We now state the results of this paper.  We set 
\[s=1/\sqrt k=\sinh 2\b J,\]
so that the low-temperature phase corresponds to $s>1$. If we define
\be D(x,y;s)=s+s\inv-(x+x\inv)/2-(y+y\inv)/2,\label{D}\ee
then we have the expansion
\be\b\inv\,\ch=1-\M^2+2\,\M^2\,\sum_{n=1}^\iy \ch^{(2n)},\label{chiexp}\ee
where 
\[\ch^{(n)}={1\ov n!}\,{1\ov(2\pi i)^n}\,\int_{\C_r}\cdots\int_{\C_r}{\prod_j x_j\inv+\prod_j y_j\inv\ov(1-\prod_jx_j)\,(1-\prod_jy_j)}\,\prod_{j<k}{x_j-x_k\ov x_j x_k-1}{y_j-y_k\ov y_j y_k-1}\,\prod_j{dx_j\,dy_j\ov D(x_j,y_j;s)}.\]
Here $\C_r$ denotes the circle with center zero and radius $r<1$ and $r$ sufficiently close to~1 (depending on $s$). All indices in the integrand run from 1 to $n$.
A derivation of this representation will be given in Appendix A.

We extend $\ch^{(n)}$ to a function of the complex variable $s$ with $|s|>1$. A {\it Nickel singularity} is a point $s^0$ on the unit circle $\T$ such that the real part of~$s^0$ is the average of the real parts of two $n$th roots of unity. 

We shall show that for~$n$ even these are the only singularities of $\ch^{(n)}$. More precisely, $\ch^{(n)}$ extends from the exterior of $\T$ to a $C^\iy$ function on $\T$ except for the Nickel singularities.\footnote{For $n$ odd our argument leaves open the possibility of other singularities. See foonote~\ref{nodd}.}

Here we use the term ``singularity'' to denote a point in no neighborhood of which a function is $C^\iy$. In the physics litearture it usually means a point beyond which a function cannot be continued analytically. It appears that $\ch^{(n)}$ satisfies a linear differential equation with only regular singular points (although the authors admit not having seen a derivation of this that they understand).\footnote{The equations for $n\le6$ have been found \cite{bhjmz}, and all their singularities are regular.} At a regular singular point the function has a series expansion whose leading term is a fractional or negative power, or a power times a power of the logarithm. Such a function cannot extend from outside $\T$ to be $C^\iy$ in a neighborhood of that point. Therefore we get the stronger result that for $n$ even $\ch^{(n)}$ extends analytically across the unit circle except at the Nickel singularities.

\bc{\bf II. Outline of the proof}\ec

With the notations 
\[F(x)={1\ov1-\prod_jx_j},\ \ F(y)={1\ov1-\prod_jy_j},\ \ 
F_{jk}(x)={1\ov1-x_j\,x_k},\ \ F_{jk}(y)={1\ov1-y_j\,y_k},\]
\[G_j(x,y;s)={1\ov D(x_j,y_j;s)},\ \ \
\D(x,y)=\Big(\prod_j x_j\inv+\prod_j y_j\inv\Big)\,\prod_{j<k}(x_j-x_k)\,(y_j-y_k),\]
all thought of as functions on $\R^n\times\R^n$, the integral becomes
\[\int_{\C_r^n}\int_{\C_r^n}F(x)\ F(y)\ 
\prod_{j<k}F_{jk}(x)\ \prod_{j<k}F_{jk}(y)\ \prod_jG_j(x,y;s)\ \D(x,y)\ dx\,dy.\]
This equals $r^{2n}$ times
\be\int_{\T^{n}}\int_{\T^{n}}F(rx)\ F(ry)\ 
\prod_{j<k}F_{jk}(rx)\ \prod_{j<k}F_{jk}(ry)\ \prod_jG_j(rx,ry;s)\ \D(rx,ry)\ dx\,dy.\label{integral}\ee

A partition of unity allows us to localize. Near any given point $(x^0,y^0)\in\T^n\times \T^n$ some of the $F$-factors may become singular as $r\to1$, and after letting $r\to1$ some of the $G$-factors may become singular as $s\to s^0\in\T$. We represent each of these potentially singular factors as an exponential integral over $\R^+$. The gradient of the exponent in the resulting integrand is approximately a linear combination with positive coefficients of certain vectors, one from each factor. Unless $s^0$ is a Nickel singularity, the convex hull of these vectors does not contain 0, a fact that  allows us to find a lower bound for the length of the gradient. (This is the crucial point in the proof.\footnote{Each of the limiting factors $F(x),\,F(y),\,F_{jk}(x),\,F_{jk}(y),\,G_j(x,y;s^0)$ may be interpreted as a distribution on $\T^n\times\T^n$. That 0 is not in the convex hull of the vectors is precisely the condition that allows one to define the product of these distributions as a distribution \cite{h}. This is what led us to the present proof.}) Then several applications of the divergence theorem give the bound $O(1)$ for the integral, uniformly in $s$ and $r$. The same is true after differentiating with respect to~$s$ any number of times. This will imply that $\ch^{(n)}$ extends to a $C^\iy$ function on $\T$ excluding these points. 

\bc{\bf III. The proof}\ec

For a given point $(x^0,y^0)=((x_j^0),(y_j^0))\in\T^n\times \T^n$ some of the factors in (\ref{integral}) become singular as $r\to 1$ and $s\to s^0$, as described above. For example $F(rx)$ becomes singular when $\prod_jx_j^0=1$ and $G_j(rx,ry;s)$ becomes singular when 
\[\Re\,x_j^0+\Re\,y_j^0=2\,\Re\,s^0.\]
There is a neighborhood of $(x^0,y^0)$ in which no other factors become singular, so that outside this neighborhood the rest of the integrand is a smooth function of $x$ and $y$ and bounded for $s$ in a neighborhood of $s^0$, together with each of its derivatives with respect to $s$. Let $\ps(x,y)$ be a $C^\iy$ function with support in this neighborhood. (Eventually the support will be taken even smaller.) We shall show that the integral (\ref{integral}), with the function $\ps(x,y)$ inserted in the integrand, is uniformly bounded for $s$ in a neighborhood of $s^0$, together with each derivatives with respect to $s$, when $r$ is taken close enough (depending on $s$) to 1.

In our neighborhood we make the variable changes
\[x_j=x_j^0\,e^{i\th_j},\ \ \  y_j=y_j^0\,e^{i\ph_j}.\]
Below we give the behavior of the reciprocals of the $F$-factors, in terms of the $\th_j,\,\ph_j$, if the factors become singular at $(x^0,y^0;s^0)$. 

\[1/F(rx)=-i\,\sum_j \th_j+O\Big((1-r)+\sum_j\th_j^2\Big),\]
\[1/F(ry)=-i\,\sum_j \ph_j+O\Big((1-r)+\sum_j\ph_j^2\Big),\]
\[1/F_{jk}(rx)=-i\,(\th_j+\th_k)+O\Big((1-r)+\th_j^2+\th_k^2\Big),\]
\[1/F_{jk}(ry)=-i\,(\ph_j+\ph_k)+O\Big((1-r)+\ph_j^2+\ph_k^2\Big).\]
We note that the real parts of these reciprocals are at least $1-r$, and so are all positive.

For any singular $G$-factors we have 
\[i/G_j(rx,ry;s)=-i\,(\a_j\,\th_j+\b_j\,\ph_j)-i\,[s+s\inv-(s^0+{s^0}\inv)]+O\Big((1-r)+\th_j^2+\ph_j^2\Big).\]
where
\[\a_j=\Im\,x_j^0,\ \ \ \b_j=\Im\,y_j^0.\]
The reason we put the factor $i$ on the left is that now the real part of the right side, which is equal to the imaginary part of the expression in brackets, is positive when $\Im\,s>0$ and $r$ is sufficiently close to 1 (depending on $s$). This we assume. (Otherwise we replace the factor $i$ by $-i$ and change signs in the definitions of $\a_j$ and $\b_j$.)

All estimates are consistent with differentiation. For example, the result of differentiating $1/F(rx)$ with respect to $\th_k$ is $-i+O((1-r)+\sum_j|\th_j|)$. 

In what follows we exclude $s^0=\pm1,\pm i$, which are Nickel singularities for even $n$. Thus we assume $(\a_j,\,\b_j)\ne(0,0)$.

Because all real parts of the reciprocals are positive they may be represented as integrals over $\R^+$. Thus, we have for any singular factor,
\[F(rx)=\int_{\R^+}e^{i\x\,(\sum_j \th_j+\,{\rm correction})}\,d\x,\]
\[F(ry)=\int_{\R^+}e^{i\e\,(\sum_j \ph_j+\,{\rm correction})}\,d\e,\]
\[F_{jk}(rx)=\int_{\R^+}e^{i\x_{jk}\,(\th_j+\th_k+\,{\rm correction})}\,d\x_{jk},\]
\[F_{jk}(ry)=\int_{\R^+}e^{i\e_{jk}\,(\ph_j+\ph_k+\,{\rm correction})}\,d\e_{jk},\]
\[G_j(x,y;s)=i\,\int_{\R^+}e^{i\z_j\,(\a_j\,\th_j+\b_j\,\ph_j+s+s\inv-s^0-{s^0}\inv+\,{\rm correction})}\,d\z_j.\]
In all of these, ``correction'' denotes $i$ times the $O$ terms above.

Thus, the integral (\ref{integral}) is replaced by one in which the cut-off function $\ps(x,y)$ is inserted into the integrand and each eventually singular factor is replaced by an integral over $\R^+$. Denote the number of these factors (and so the number of $(\x,\,\e,\,\z)$-integrations) by $N$. We change the order of integration and integrate first with respect to the $\th_j,\,\ph_j$. We want to apply the divergence theorem so that we eventually get a bound $O(R^{-N-1})$, where $R$ is the radial variable in the $N$-dimensional $(\x,\,\e,\,\z)$-space. To do this we have to find a lower bound for the length of the gradient of the sum of the exponents coming from the $(\x,\,\e,\,\z)$-integrations. 

We define the following vectors in $\R^n\times\R^n$:
\[\begin{array}{cc}
X=(1\ 1\ \cd\ 1\ 0\ 0\ \cd\ 0)\\&\\
Y=(0\ 0\ \cd\ 0\ 1\ 1\ \cd\ 1)\\&\\
X_{jk}=(0\cd 1\cd 1\cd\ 0\ 0\ \cd)\\&\\
Y_{jk}=(\cd\ 0\ 0\ \cd 0\ 1\cd 1\cd)\\&\\
Z_j=(0\ \cd0\ \a_j\ 0\cd0\ \b_j\cd 0).
\end{array}\]
Let us explain. The first $n$ components are the $\th_j$ components, the last $n$ the $\ph_j$ components. For $X$ the ones are the first $n$ components and the zeros are the rest, and for $Y$ these are reversed. For $X_{jk}$ the ones are components $j$ and $k$ and the others are zero, and for $Y_{jk}$ the ones are components $n+j$ and $n+k$ and the others are zero. For $Z_j$ component $j$ is $\a_j$ and component $n+j$ is $\b_j$, and the others are zero.

Aside from the factor $i$ and the correction term from each summand, the gradient of the sum of the exponents is the subsum of 
\be\x\,X+\e\,Y+\sum_{j<k}\x_{jk}\,X_{jk}+\sum_{j<k}\e_{jk}\,Y_{jk}+\sum_j \z_j\,Z_j\label{sum}\ee
containing the $N$ $(\x,\,\e,\,\z)$-variables that actually appear.

\noi{\bf Lemma 1}. Suppose that $n$ is even and that $s^0$ is not a Nickel singularity. Then 0 is not in the convex hull of those of the vectors $X,\,Y,\,X_{jk},\,Y_{jk},\,Z_j$ that appear in the subsum of (\ref{sum}).

\noi{\bf Proof}. We show that if a linear combination of these vectors with nonnegative coefficients is zero, but not all the coefficients are zero, then $s^0$ is a Nickel singularity. We say that a vector ``appears'' in the linear combination if its coefficient is nonzero. Some $Z_j$ must appear since all the others have nonnegative components and at least one positive component. (Recall that $Z_j$ appears when $\Re\,x_j^0+\Re\,y_j^0=2\,\Re\,s^0$.)

If $X_{jk}$ appears then then so must $Z_j$ and $Z_k$ and $\a_j,\,\a_k<0$, to cancel the nonzero components of $X_{jk}$. But $X_{jk}$ appears only when $x_j^0\,x_k^0=1$, so $\a_j+\a_k=0$, which is a contradiction. Thus no $X_{jk}$ appears. Similarly no $Y_{jk}$ appears.

Since some $Z_j$ appears either $X$ or $Y$ must. Suppose that $X$ appears. (In particular $\prod x_j^0=1$.)  Then all $\a_j<0$, and if the coefficient of $X$ is $c_X$ the  coefficient of $Z_j$ must be $-c_X/\a_j$. 
 
There are two subcases:

\noi(i) $Y$ appears: (In particular $\prod y_j^0=1$.) In analogy with the above, if the coefficient of $Y$ is $c_Y$ then the coefficient of $Z_j$ is $-c_Y/\b_j$. Thus $\a_j/\b_j=c_X/c_Y$ for all $j$. We claim that this implies that all $x_j$ are equal and all $y_j$ are equal. Consider pairs $(x,y)$ with both in the lower half-plane, and $\Re\,x+\Re\,y=2\,\Re\,s^0$. Set $x=e^{i\th},\ y=e^{i\ph}$. It is an exercise in calculus to show that as $\th$ increases while $\cos\,\th+\cos\,\ph$ remains constant the ratio $\Im\,x/\Im\,y=\sin\th/\sin\,\ph$ strictly decreases if $\Re\,s^0>0$ and strictly increases if $\Re\,s^0<0$. Therefore this ratio determines $\th$, and so $x$. Similarly the ratio determines $y$. So all $x_j$ are equal and all $y_j$ are equal, as claimed. They must both be $n$th roots of unity, so $s^0$ is a Nickel singularity.

\noi(ii) $Y$ does not appear: Since all $Z_j$ appear, we must have all $\b_j=0$ in this case. So all $y_j^0=\pm1$. If some $y_j^0=1$ then $\Re\,s^0>0$, because if $\Re\,s^0$ were negative it could not be the average of 1 and some $\Re\,x_j^0$. Then all $y_j^0=1$, for the same reason. Hence each $\Re\,x_j=2\,\Re\,s-1$, and since all $\a_j<0$ this implies that all $x_j$ are equal, and equal to some $n$th root of unity.  Thus $s^0$ is a Nickel singularity. If some $y_j^0=-1$, and therefore all $y_j=-1$, this again implies that all $x_j^0$ equal some $n$th root of unity. Since $n$ is even $s^0$ is again a Nickel singularity.\footnote{Since $-1$ is not an $n$th root of unity when $n$ is odd, these $s^0$ are not Nickel singularities.\label{nodd}}\qed

If 0 is not in the convex hull of vectors then there is a lower bound for linear combinations of them with nonnegative coefficients, even when the vectors are perturbed.

\noi{\bf Lemma 2.} Assume 0 is not in the convex hull of the vectors $V_1,\ldots,V_N$. Then for sufficiently small $\ep>0$ there is a $\d>0$ such that, for vectors $U_j$ with $|U_j-V_j|<\ep$ and coefficients $c_j\ge0$, we have 
\be\Big|\sum_j c_j\,U_j\Big|\ge \d\,\sum_j c_j.\label{ineq}\ee

\noi{\bf Proof}. Suppose the result is not true. Then there is a sequence $\ep_k\to0$, vectors $U_{j,k}$ with $|U_{j,k}-V_j|\le \ep_k$, and coefficients $c_{j,k}\ge0$ such that for each $k$,
\[\Big|\sum_j c_{j,k}\,U_j\Big|<{1\ov k}\sum_j c_{j,k}.\]
By homogeneity we may assume that each $\sum_j c_{j,k}=1$. Then, by taking  subsequences, we may assume that each $c_{j,k}$ converges as $k\to\iy$ to some $c_j$. Then $\sum_j c_j=1$, and each $U_{j,k}\to V_j$, so $\sum_j c_j\,V_j=0$. This is a contradiction.\qed

\noi{\bf Lemma 3.} Assume $n$ is even and $s^0$ is not a Nickel singularity. There is a neighborhood of $(x^0,y^0)$ such that if $\ps(x,y)$ is a $C^\iy$ function with support in that neighborhood then the integral (\ref{integral}), with $\ps$ inserted in the integrand and $r$ sufficiently close to~1 (depending on $s$), is bounded in a neighborhood of $s=s^0$; and the same is true for each derivative with respect to $s$.

\noi{\bf Proof}. With the same $\ps$ as above, we combine Lemmas 1 and 2 to deduce that if $r$ is close enough to 1, and the support of $\psi$ is small enough, then the length of the gradient of the exponent in the integral is at least a constant times the sum of the coefficients in the subsum of (\ref{sum}) that arises. Therefore $N+1$ applications of the divergence theorem shows that the integral over the $\th_j,\,\ph_j$ has absolute value at most a constant times $1/R^{N+1}$, where $R$ is the radial variable in the $N$-dimensional $(\x,\,\e,\,\z)$-space.\footnote{We explain this in Appendix B.}  Therefore the integral (\ref{integral}) with $\ps(x,y)$ inserted in the integrand, which results after integration over the $(\x,\,\e,\,\z)$, is $O(1)$ uniformly for $s$ in  a neighborhood of $s^0$. (The integral over $R<1$ is clearly bounded.) Differentiating with respect to~$s$ any number of times just brings down powers of the $\z_j$, and so only requires more applications of the divergence theorem.\qed
\sp

\noi{\bf THEOREM}. When $n$ is even $\ch^{(n)}$ extends to a $C^\iy$ function on $\T$ except at the Nickel singularities.
\sp
 
\noi{\bf Proof}. Assume $s^0$ is not a Nickel singularity. Each $(x^0,y^0)$ has a neighborhood given by Lemma 3. Finitely many of these neighborhoods,  cover $\T^n\times\T^n$. We can find a $C^\iy$ partition of unity 
$\{\ps_i(x,y)\}$ such that the support of each $\ps_i$ is contained in one of these neighborhoods. Each integral (\ref{integral}) with $\ps_i(x,y)$ inserted in the integrand and $r$ sufficiently close to 1, together with each derivative with respect to $s$, is bounded in a neighborhood of $s=s^0$. Therefore the same is true of (\ref{integral}) itself, and therefore for $r^n$ times (\ref{integral}), which is independent of $r$, and therefore for $\ch^{(n)}$. This implies that $\ch^{(n)}$ extends to a $C^\iy$ function on $\T$ in a neighborhood of $s^0$.\footnote{We explain this in Appendix C.}\qed
\pagebreak

\bc{\bf Appendix A}\ec

For $T<T_c$ and $N\ge0$ we have the following Fredholm determinant representation of the spin-spin correlation function (see \cite[p. 375]{PT} or \cite[p. 142]{palmer}):
\[\la \s_{00}\,\s_{MN}\ra =\M^2 \det(I+g_{MN}).\]
The operator has kernel
\[g_{MN}(\th_1,\th_2) = e^{i M\th_1-N\g(e^{i\th_1})}\,h(\th_1,\th_2),\]
where
\[h(\th_1,\th_2)={\sinh\hf(\g(e^{i\th_1})-\g(e^{i\th_2}))\ov\sin\hf(\th_1+\th_2)},\]
and $\g(z)$ is defined by
\[\cosh\g(z)= s+s\inv-(z+z\inv)/2,\]
with the condition that $\g(z)$ is real and positive for $|z|=1$.
The operator acts on $L^2(-\pi,\pi)$ with weight function
\[{1\ov2\pi\,\sinh \g(e^{i\th})}.\]

Using the identity (see \cite[(5.5)]{PT} or \cite[(2.69)]{palmer})
\[\det\left(h(\th_j,\th_k)\right)=\prod_{j<k} [h(\th_j,\th_k)]^2, \]
and the Fredholm expansion we obtain that $\langle \s_{00}\,\s_{MN}\rangle$ equals 
\be\M^2\,\sum_{n=0}^\iy {1\ov(2n)!} {1\ov (2\pi)^{2n}}\int_{-\pi}^\pi\cdots \int_{-\pi}^\pi
\prod_{j<k} [h(\th_j,\th_k)]^2\, \prod_j e^{i M\th_j-N\g(e^{i\th_j})} 
{d\th_j\ov\sinh\g(e^{i\th_j})}.\label{correxp}\ee
Here all indices run from 1 to $2n$. We used the fact that since the matrix $(h(\th_j,\th_k))$ is antisymmetric its odd-order determinants vanish.

We have the identity, observed in \cite{nickel2},
\[{\sinh(\hf(\g(e^{i\th_1})-\g(e^{i\th_2}))\ov\sin(\hf(\th_1+\th_2))}= 
 {\sin(\hf(\th_1-\th_2))\ov\sinh(\hf(\g(e^{i\th_1})+\g(e^{i\th_2}))}\]
Therefore, with $x_j=e^{i\th_j}$,
\[[h(\th_1,\th_2)]^2= {e^{-\g(x_1)}-e^{-\g(x_2)}\ov1-e^{-\g(x_1)-\g(x_2)}} \, {x_1-x_2\ov1-x_1 x_2}.\]

With $D(x,y;s)$ defined by (\ref{D}) a short calculation shows that
\[y\,D(x,y;s)=-\hf(y-e^{-\g(x)})\,(y-e^{\g(x)}).\]
Thus inside the unit circle $1/(y\,D(x,y;s))$ has a pole at $y=e^{-\g(x)}$ with residue $1/\sinh\g(x)$. It follows that
\[{1\ov(2\pi i)^{2n}}\,\int_{\C_r}\cdots\int_{\C_r}\prod_{j<k}{y_j-y_k\ov1-y_jy_k}\,
\prod_j{y_j^{N-1}\,dy_j\ov D(x_j,y_j;s)}=
\prod_j{e^{-N\g(x_j)}\ov\sinh\g(x_j)}\,\prod_{j<k}{e^{-\g(x_j)}-e^{-\g(x_k)}\ov1-e^{-\g(x_j)-\g(x_k)}}.\] We deduce that the integral in (\ref{correxp}) equals 
\be{1\ov(2\pi)^{2n}}\int_{\C_r}\cdots\int_{\C_r}\prod_{j<k}{y_j-y_k\ov1-y_jy_k}\,{x_j-x_k\ov1-x_jx_k}\,\prod_j{x_j^{M}\,y_j^{N}\ov D(x_j,y_j;s)}\,\prod_j{dx_j\ov x_j}\,{dy_j\ov y_j}.\label{corrint}\ee

It remains to compute
\[\sum_{M,N\in\Z}\left\{\la \s_{0,0}\,\s_{M,N} \ra-\M^2 \right\}.\] 
Since $\la \s_{0,0}^{\,2}\ra=1$ the $(0,0)$ term equals $1-\M^2$. For the remaining terms, subtracting $\M^2$ in the summand is the same as taking the sum in (\ref{correxp}) only over $n>0$.

To compute the sum over $(M,N)\ne(0,0)$ we use the fact that $\la \s_{0,0}\,\s_{M,N} \ra$ is even in $M$ and in $N$, so
\[\sum_{(M,N)\ne(0,0)}=4\,\sum_{M,N\ge 0}-2\,\sum_{M=0,N\ge 0}-2\,\sum_{N=0,M\ge 0}\] 
and find that after summing, the factor $\prod_j x_j^{M}\,y_j^{N}$ in the integrand in (\ref{corrint}) gets replaced~by
\[2\,{\prod_j x_j+\prod_j y_j\ov(1-\prod_j x_j)\,(1-\prod_j y_j)}.\]
This gives (\ref{chiexp}).

\bc{\bf Appendix B}\ec

Suppose $f$ and $g$ are $C^\iy$ functions on $\R^d$, with $f$ having compact support, and we have an integral
\[\int f(\th)\,e^{g(\th)}\,d\th.\]
We write it as
\[\int f(\th){\Del g(\th)\ov|\Del g(\th)|^2}\,\cdot\, \Del e^{g(\th)}\,d\th.\]
If define the operator $L$ by
\[(Lf)(\th)=-\Del\,\cdot\,f(\th){\Del g(\th)\ov|\Del g(\th)|^2},\]
then $q$ applications of the divergence theorem show that the integral equals
\[\int (L^qf)(\th)\;e^{g(\th)}\;d\th.\]

Now we have

\noi(a) $L^q f$ is a linear combination of (partial) derivatives of $f$ with coefficients that are homogeneous polynomials of degree $q$ in derivatives of the components of $\Del g/|\Del g|^2$; 

\noi(b) each $p$th derivative of each component of $\Del g/|\Del g|^2$ equals 
$1/|\Del g|^{2p+2}$ times a homogeneous polynomial of degree $2p+1$ in derivatives of $g$.

Assume that we also have

\noi(c) $|\Del g(\th)|\ge\m$ and each derivative of $g(\th)$ is $O(\m)$;

\noi(d) each derivative of $f(\th)$ is $O(1)$.

Then assuming that $\Re\,g$ is uniformly bounded above, we can conclude that
\[\int_{\R^d} f(\th)\,e^{g(\th)}\,d\th=O(\m^{-q})\ \  \textrm{for all}\ q.\]

In the application in Lemma 3 we have $d=2n$, $g$ is the sum of the exponents in the integrals, $f$ is the product of other integrands, and $\m$ can be taken to be a small constant times the sum of the coefficients in the subsum of (\ref{sum}).

\bc{\bf Appendix C}\ec

Suppose $\U$ is an open set in $\T$, that $f$ is analytic in the region
\[\O=\{Rs:s\in \U,\ \ 1<R<1+\d\},\]
and that $f$ and each of its derivatives is bounded in $\O$. 
We show that $f$ extends to a $C^\iy$ function on $\O\cup\U$. 

Pick any $s_0\in\O$. We have for each $k\ge0$ and $s'\in\O$,
\[ f^{(k)}(s')=f^{(k)}(s_0)+\int_{s_0}^{s'} f^{(k+1)}(t)\,dt.\]
Since $f^{(k+1)}$ is bounded, this shows that that $f^{(k)}$ extends continuously to $\O\cup \U$. Denote by $f_k(s)$ this extension. In paticular $f_0$ is the continuous extension of $f$. We show that it belongs to $C^\iy$.

We show by induction that $f_0\in C^k$. We know this for $k=0$. Assuming this for~$k$, we see that for $s\in \U$,
\[{d^k\ov ds^k}f_0(s)=\lim_{s'\to s}{d^k\ov ds'{^k}} f(s')=
f^{(k)}(s_0)+\int_{s_0}^s f_{k+1}(t)\,dt.\]
It follows that $f_0$ is $k+1$ times differentiable and
\[{d^{k+1}\ov ds^{k+1}}f_0(s)=f_{k+1}(s)=
\lim_{s'\to s} {d^{k+1}\ov ds'{^{k+1}}}f(s').\]
This gives the assertion.

\bc{\bf Acknowledgments}\ec

That authors thank Tony Guttmann, Masaki Kashiwara, Jean-Marie Maillard, Bernie Nickel, Jacques Perk, and, especially, Barry McCoy for helpful communications.

This work was supported by National Science Foundation grants DMS--1207995 (first author) and DMS--0854934 (second author).

\end{document}